\documentclass[conference]{IEEEtran}
\usepackage{cite}
\usepackage{amsmath,amssymb,amsfonts}
\usepackage{algorithmic}
\usepackage{enumitem}
\usepackage{graphicx}
\usepackage{textcomp}
\usepackage{booktabs}
\usepackage{siunitx}
\usepackage{caption}
\usepackage{placeins}
\usepackage{dirtytalk}
\usepackage{xcolor}
\usepackage{pifont}
\usepackage{tikz}
\usepackage{scalefnt}
\usepackage{balance}
\usepackage[hidelinks]{hyperref}
\usepackage{orcidlink}
\usepackage{anyfontsize}
\def\BibTeX{{\rm B\kern-.05em{\sc i\kern-.025em b}\kern-.08em
    T\kern-.1667em\lower.7ex\hbox{E}\kern-.125emX}}

\usepackage{changepage}

 \usepackage{framed}
 \makeatletter
  {\par\unskip\endMakeFramed}
 \makeatother
 
 \definecolor{textboxshade}{rgb}{0.93,0.93,0.93}
 
 \definecolor{darkblue}{rgb}{0.2, 0.2, 0.2}
 
 \newenvironment{textbox}{%
   \def\FrameCommand{%
     \hspace{1pt}%
     {\color{darkblue}\vrule width 2pt}%
     {\color{textboxshade}\vrule width 4pt}%
     \colorbox{textboxshade}%
   }%
   \MakeFramed{\advance\hsize-\width\FrameRestore}%
   \noindent\hspace{-1pt}
   \begin{adjustwidth}{}{7pt}%
   \vspace{2pt}\vspace{2pt}%
 }
 {%
   \vspace{3pt}\end{adjustwidth}\endMakeFramed%
 }
 
 \newcounter{resultcounter}

 \newcounter{patterncounter}
\makeatletter
\def\namedlabel#1#2#3{\begingroup
    #3%
    \def\@currentlabel{\textbf{#2}}%
    \phantomsection\label{#1}\endgroup
}
\makeatother

\newcommand{\approach}{ToMMY}

\newcommand{\captiondesc}[2]{\caption[#1]{#1 #2}}

\newcommand*\figrectangle[1]{\tikz[baseline=(char.base)]{
            \node[shape=rectangle,draw,inner sep=2pt] (char) {#1};}}

\begin{document}

\title{
    What You Need is What You Get: Theory of Mind for an LLM-Based Code Understanding Assistant
} 

\author{\IEEEauthorblockN{Jonan Richards \orcidlink{0009-0007-1218-8599}, Mairieli Wessel \orcidlink{0000-0001-8619-726X}}
\IEEEauthorblockA{\textit{Radboud University} \\
Nijmegen, The Netherlands\\
jonan.richards@ru.nl, mairieli.wessel@ru.nl}
}


\maketitle

\begin{abstract}
A growing number of tools have used Large Language Models (LLMs) to support developers' code understanding. However, developers still face several barriers to using such tools, including challenges in describing their intent in natural language, interpreting the tool outcome, and refining an effective prompt to obtain useful information. In this study, we designed an LLM-based conversational assistant that provides a personalized interaction based on inferred user mental state (e.g., background knowledge and experience). We evaluate the approach in a within-subject study with fourteen novices to capture their perceptions and preferences. Our results provide insights for researchers and tool builders who want to create or improve LLM-based conversational assistants to support novices in code understanding.
\end{abstract}


\section{Introduction}




Large Language Models (LLMs) have shown promising results supporting code comprehension tasks and are often perceived as useful by developers~\cite{ross2023a,khojah2024,nam2024}. However, the extent to which users benefit from these tools depends on their background knowledge and experience levels~\cite{pinto2024a,nam2024}. Developers often experience difficulty in writing prompts that are effective at addressing their information needs~\cite{nam2024,khojah2024}, particularly novices~\cite{nguyen2024}. Many studies suggest the potential of AI assistants learning from users to create personalized guidance~\cite{hassan2024,ross2023a,pinto2024a}. In fact, successful explanations should fill the gaps in questioners' background knowledge~\cite{faye1999}, while using language that is understandable to the questioner~\cite{walton2004}. Combined, these findings indicate a need for LLMs to infer and adapt to developers' needs, intents, knowledge, experience, and preferences.

Humans' ability to reason about what others are thinking is called Theory of Mind (ToM). LLMs have performed well on a range of ToM tasks~\cite{moghaddam2023,zhou2023,zhu2024}. Whether this is due to LLMs having an innate ToM capacity~\cite{jamali2023} or their ability to find shallow heuristics to perform well~\cite{shapira2023a} remains an open debate. Still, by reflecting on users' minds, LLMs can produce insightful information in practical settings~\cite{leer2023}.

We hypothesize that LLMs can gather a substantial amount of useful ToM information from interactions between a developer trying to understand unfamiliar code, and a conversational agent. This led us to create a conversational agent called \textbf{\approach{}} (\textbf{T}heory \textbf{o}f \textbf{M}ind: \textbf{M}aking explanations \textbf{Y}ours). \approach{} comprises a chain of prompts meant to generate insights about a developer's mental state. It then uses these insights to provide personalized explanations about code. 
In this paper, we evaluate how \approach{} compares to a more basic agent. Our work is the first application and evaluation of ToM prompting techniques in LLMs supporting Software Engineering activities. This study focuses on the following research questions:
\begin{description}
    \item[\namedlabel{rq:1}{RQ1}{RQ1}] How do novices interact with LLM-based assistants when trying to understand unfamiliar code?

    \item[\namedlabel{rq:2}{RQ2}{RQ2}] How does using \approach{} impact novices' code understanding?

    \item[\namedlabel{rq:3}{RQ3}{RQ3}] How do novices perceive interacting with LLM-based assistants to understand unfamiliar code?
\end{description}

To answer these research questions, we conducted a within-subject study with 14 novices, including undergraduate and graduate students. Participants interacted with each approach (\approach{} vs. simple conversational agent) individually and comparatively. Our findings reveal that novices exhibited distinct interaction styles based on whether they phrased some questions as hypotheses (or not). While interacting with \approach{}, participants less frequently stated their intent or asked follow-up questions, and slightly more often provided instructions regarding the response format. Also, using \approach{} had distinct impacts on novices' code understanding depending on their interaction styles.
We made our supplementary material available for replication purposes~\cite{replication}.

\section{Related Work}
Researchers have investigated how developers use and interact with LLMs for development-related tasks \cite{khojah2024,Liang2024,mozannar2024,ross2023,ross2023a}. Ross et al. \cite{ross2023,ross2023a} suggests adapting programming assistants to the individual strengths and needs of its users toward providing personalized responses. 
A growing number of tools have also recently brought LLMs into the code editor to support code understanding~\cite{liffiton2023codehelp,nam2024,yan2024ivie}.
Nam et al.~\cite{nam2024}, for instance, proposed an approach that supports on-demand explanations of code. 
Complementing previous literature, we focused on personalizing developers' interactions with conversational LLM-based code understanding assistants using existing Theory of Mind prompting techniques.

\section{\approach{}'s design}
The design of \approach{} was motivated by findings from previous studies of LLMs, which found that predicting and reflecting on mental states improve LLMs' performance~\cite{zhou2023,leer2023}. Furthermore, perspective-taking is an effective way to enhance the ToM performance of LLMs~\cite{wilf2023,zhu2024}. Wilf et al.\cite{wilf2023} highlights the need to separate mental state inference and question answering into two prompts. The prompt engineering process was conducted through manual evaluation, in addition to a pilot study used to evaluate the last iteration of prompts.

The first of \approach{}'s \textit{three prompts} identifies aspects of the user's mental state relevant to producing a response, and phrases these as open questions about the user. Not having this step, and directly prompting for the user's mental state, was found to produce irrelevant mental state data. In the second prompt, we instruct the LLM to take the perspective of the user to answer these questions, based on the conversation history. These answers constitute the inferred mental state. Finally, in the third prompt, the LLM is instructed to respond to the user, considering the mental state.

Each of \approach{}'s prompts includes context for the LLM, containing a code snippet, the conversation history, and the user's latest input. The basic agent, used as a control condition in the user study, comprises a single prompt containing the same context, and an instruction to respond to the user.

The LLM we used in our evaluation was GPT-3.5, a common choice in research on LLMs for Software Engineering~\cite{hou2024}. We set the model's temperature, a hyper-parameter related to output randomness, to 0. This helped isolate the impact of using different prompts. We limited the history to the 10 most recent interactions, and capped the length of input messages to 1,000 characters, to prevent the prompt from overflowing the LLM's token limit. A pilot study exposed no issues with this strategy.

\section{Research Design}
When preparing and executing this study, we followed the guidelines set by Radboud's Research Ethics Committee and received approval from that committee to conduct our study.

\noindent\textbf{Pilot sessions.}
We conducted pilot sessions with three novice programmers: one PhD student, one master's student, and one third-year bachelor's student. The participants suggested a few minor adjustments, which were incorporated into \approach{} and the experiment instruments. Due to the participant's fatigue, we time-boxed the code understanding tasks (20 minutes) and quizzes (10 minutes). Data from pilot sessions were discarded.

\noindent\textbf{Participant recruitment.}
We used a convenience sampling strategy to recruit participants for our experiment sessions, inviting students through (i) short presentations at lectures, (ii) study programme-wide messaging groups, and (iii) outreach to personal contacts. The study was not part of any course; all participants were volunteers and signed a consent form before their sessions. The only prerequisite was a low-level programming experience (e.g., having taken a programming-related course or having coded as a hobby). In total, 6 undergraduate and 8 graduate students were recruited. Their experience with programming ranged between low and experienced (with an average 5.7 on a scale from 1 to 10). Participants also reported having low (2), moderate (7), considerable (2), and extensive (1) familiarity with LLMs.

        






\noindent\textbf{Experimental sessions.}
We conducted a series of synchronous within-subject sessions with participants using both \approach{} (\textit{treatment}) and the basic agent (\textit{control}). Using variance minimization~\cite{sella2021}, participants were sorted into two groups balanced in terms of participants' programming experience, LLM familiarity, and background knowledge regarding natural language processing and data analysis. This grouping was used to determine the order in which the two agents were encountered, mitigating order effects.

Before the session, participants received email instructions, a survey with demographic questions, and a consent form. Participants could choose between online or in-person sessions. Each session started with a brief explanation of the research objectives and guidelines. Then, participants were asked to complete two code comprehension tasks (one with each conversational agent). Participants were presented with a different code snippet for each task and tasked with understanding it as thoroughly as possible within a 20-minute time limit. 

We induced the participant's need to interact with the agents by choosing code snippets that require domain knowledge to understand (i.e., data analysis and natural language processing). The two code snippets written in Python were carefully selected from open-source repositories on GitHub and not used for prompting engineering. Code comments were removed to increase the need for understanding, but typos or other implementation choices were left.

After each task, to measure their code understanding we asked participants to complete a few closed-ended questions inspired by previous literature~\cite{izu2019} (max. 10 minutes). We also asked them to complete a mid-term survey (right after the first task) and a post-study survey at the end of the experiment to elicit their perceptions of the approaches. 

\begin{figure}[!htp]
\centering
    \includegraphics[width=\linewidth]{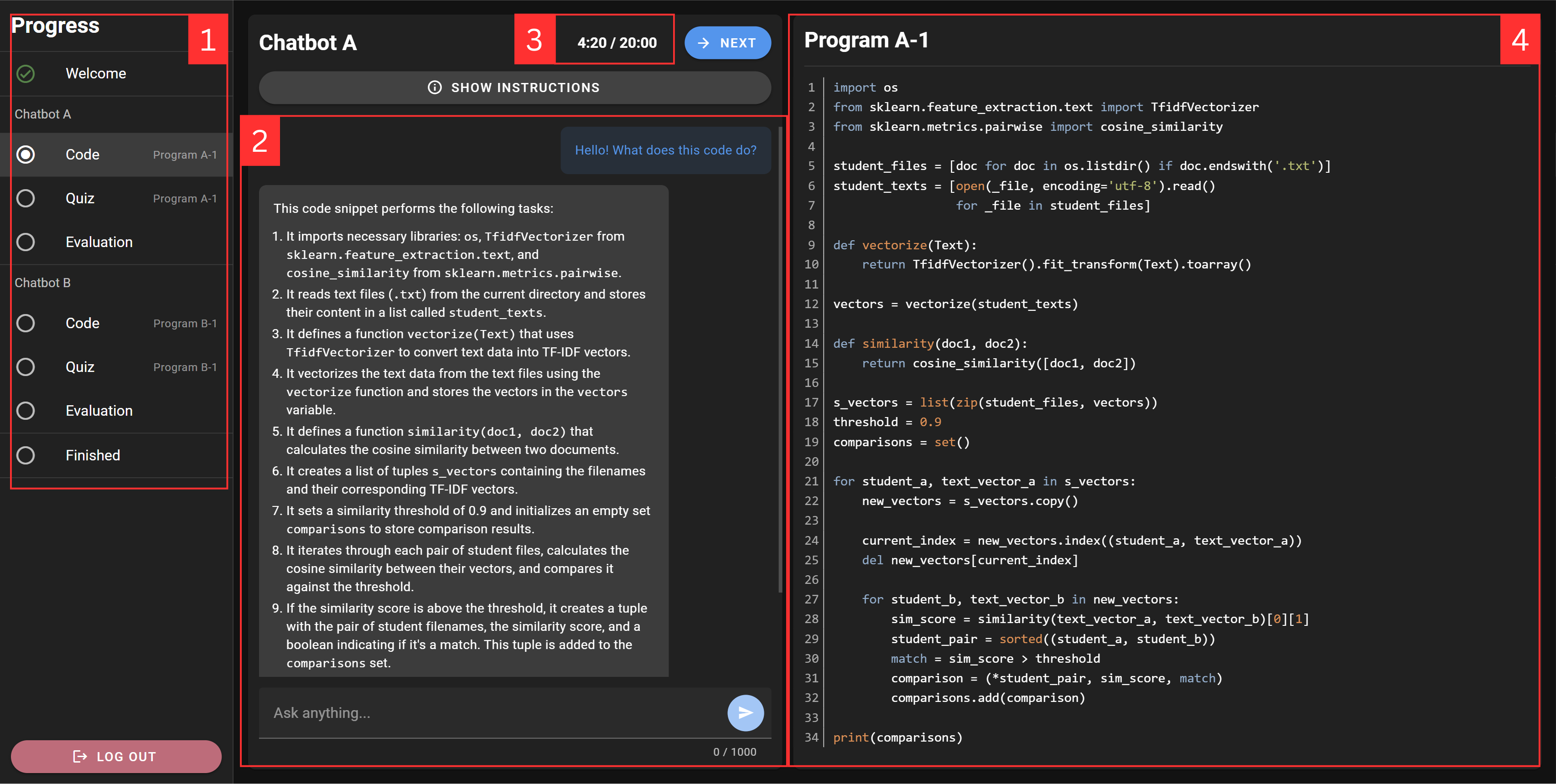}
    \caption{Web-based interface designed for the experiment}
    \label{fig:interaction}
\end{figure}

\noindent\textbf{Web-based evaluation interface.}
To enable the user study, we developed a custom-built web-based tool. This tool included a chat interface (Figure~\ref{fig:interaction}-\figrectangle{2}) for interacting with the conversational agents, alongside a code snippet (Figure~\ref{fig:interaction}-\figrectangle{4}) to mimic the experience of being in a code editor. It also displayed instructions and allowed participants to navigate the study steps (i.e., tasks, quiz questions, and evaluation forms). Participants could track their progress in the left-side menu (Figure~\ref{fig:interaction}-\figrectangle{1}). A timer for each step was also displayed (Figure~\ref{fig:interaction}-\figrectangle{3}).

\noindent\textbf{Data analysis.}
We analyzed participants' interactions with the agents (\ref{rq:1}) using \emph{open} card sorting~\cite{zimmermann2016card}. First, the first two authors independently analyzed the inputs for the first two participants (cards) and applied codes, sorting them into meaningful groups. A discussion meeting followed this step until a consensus was reached on the categorization. The first author then coded the remainder of the participant interaction data. The researchers analyzed the categories to refine the classification and group-related codes into higher-level themes.
 
To answer \ref{rq:2}, we used linear regression to assess each approach's impact, considering the effect of participants' programming experience, background experience, and familiarity with LLM tools. We chose these three metrics as our independent variables and a variable to indicate whether a participant used our approach for the task. To measure code understanding, we considered the number of correct answers on the quizzes and the time participants took to complete the task. We fit two separate regression models to these two dependent variables. 
After observing distinct patterns during the qualitative analysis of participants' interactions, we performed additional analysis for code understanding. We fit two more regression models on the quiz scores: one for participants who interacted by phrasing hypotheses and the other for participants who did not.

In \ref{rq:3}, we assessed the perceived usefulness, ease of use, and cognitive load of working with each agent by applying the Technology Acceptance Model (TAM)~\cite{davis1989perceived} and Task Load Index (NASA-TLX)~\cite{hart1988development}, respectively. Additionally, we quantitatively analyzed the open feedback collected after completing each task. This was processed using card sorting, similar to \ref{rq:1}.
\label{section:user_study}

\section{Findings}
\subsection{Interaction (RQ1)}
\label{section:results_interaction}
Participants did not interact more frequently, or for a longer time, with either agent (\autoref{table:interaction_stats_models}). This was determined using two-sided paired t-tests (Shapiro-Wilk tests showed normality of differences for these variables, $p=0.88$ and $0.43$ respectively).

\begin{table}[htbp]
    \scalefont{0.8}
    \centering
    \begin{tabular}{l *{2}{S[table-format=4.1, round-mode=places, round-precision=2] S[table-format=3.1, round-mode=places, round-precision=2]} S[table-format=1.2, round-mode=places, round-precision=2]} 
        \toprule
        & \multicolumn{2}{c}{Control}
        & \multicolumn{2}{c}{\approach{}}
        & \multicolumn{1}{c}{} \\

        \cmidrule(lr){2-3}\cmidrule(lr){4-5}

        & \multicolumn{1}{c}{mean} 
        & \multicolumn{1}{c}{std}
        & \multicolumn{1}{c}{mean} 
        & \multicolumn{1}{c}{std} 
        & \multicolumn{1}{c}{$p$-value} \\

        \midrule
        
        \# of interactions
        & 7.214286 & 4.560340
        & 7.142857 & 4.704452 
        & 0.9579201278402341 \\
        
        Interaction time (s)
        & 739.000000 & 363.664258
        & 717.142857 & 319.379894
        & 0.7242963106617961 \\

        \bottomrule
    \end{tabular}

    \caption{Average number of interactions and interaction time per participant for each task, broken down by model.}
    \label{table:interaction_stats_models}
\end{table}

In our qualitative analysis of the participants' inputs, we grouped the codes into 5 categories, which are described in the following sections.

\begin{figure}[htbp]
    \centering
    \includegraphics[width=0.8\linewidth]{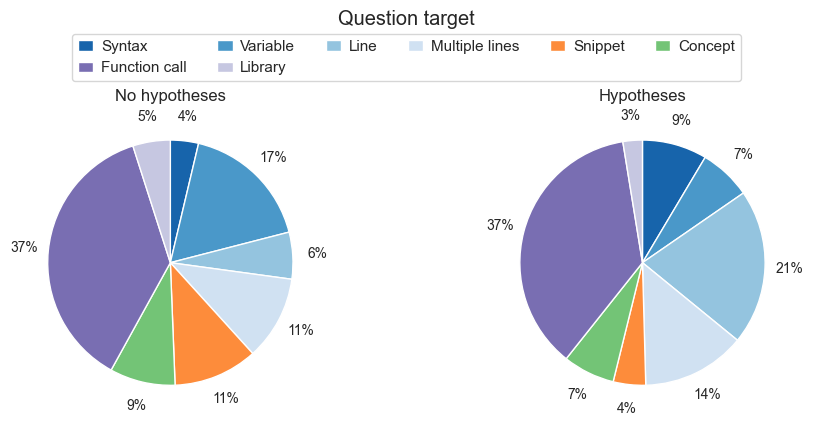}
    \caption{Distribution of question targets, broken down by participants who did not (left) and did (right) use hypotheses.}
    \label{figure:interaction_target_hypotheses}
\end{figure}

\noindent\textbf{Question.}
All but four of the 201 recorded interactions contained a question, and only one input contained two questions. These question were phrased either as an open question (\say{What does ... do?}), a hypothesis (\say{Does ... do ...?}), an instruction (\say{Explain ...}), or implicitly (e.g. by only providing a line of code for the agent to explain). We clustered participants based on if they phrased at least one question as a hypothesis, and found that 6 out of 14 participants did. We found that this cluster had significantly more interactions and spent a significantly longer time interacting with the agents than the participants who did not use hypotheses (using two-sided Mann-Whitney U tests, \autoref{figure:interaction_target_hypotheses}). The difference in programming experience and LLM familiarity between these clusters of participants was negligible.

\begin{table}[htbp]
    \scalefont{0.8}
    \centering
    \begin{tabular}{l *{2}{S[table-format=4.1, round-mode=places, round-precision=2] S[table-format=3.1, round-mode=places, round-precision=2]} S[table-format=1.2, table-space-text-post=\textsuperscript{***}, round-mode=places, round-precision=2]} 
        \toprule
        & \multicolumn{2}{c}{No hypotheses}
        & \multicolumn{2}{c}{Hypotheses}
        & \multicolumn{1}{c}{} \\

        \cmidrule(lr){2-3}\cmidrule(lr){4-5}

        & \multicolumn{1}{c}{mean} 
        & \multicolumn{1}{c}{std}
        & \multicolumn{1}{c}{mean} 
        & \multicolumn{1}{c}{std} 
        & \multicolumn{1}{c}{$p$-value} \\

        \midrule

        Programming exp.
        & 4.812500 & 1.944544
        & 4.666667 & 1.834848
        & 0.9479582550113652 \\

        LLM familiarity
        & 2.875000 & 0.834523
        & 3.166667 & 0.983192
        & 0.6749188570622133 \\
        
        \# of interactions
        & 10.250000 & 6.860862
        & 19.833333 & 5.492419
        & 0.01986361483169752\textsuperscript{*}\\
        
        Interaction time (s)
        & 1072.25 & 490.628387
        & 1968.00 & 443.946844
        & 0.007992007992007992\textsuperscript{**}\\

        \bottomrule
            
        \multicolumn{6}{r}{\textsuperscript{*}$p < 0.05$, \textsuperscript{**}$p < 0.01$, \textsuperscript{***}$p < 0.001$}
    \end{tabular}

    \caption{Participants' background experience and interaction statistics, broken down by participants who did not (left) and did (right) use hypotheses.}
    \label{table:interaction_stats_hypotheses}
\end{table}

\noindent\textbf{Question target.} 
This can be either a fine-grained code feature (such as syntax, a variable, a line of code or multiple lines of code); the entire snippet; an external code feature (such as an external function call or a library); or a programming concept. \autoref{figure:interaction_target_hypotheses} shows that participants who did not phrase questions as hypotheses tended to ask more questions about the entire snippet, compared to participants who did employ hypotheses. Instead, the latter more often targeted their questions at fine-grained code features. There were no discernible patterns regarding which agent was used.

\noindent\textbf{Question intent.} 
Most questions did not specify what aspect of a code feature they wanted the agent to explain (e.g. in \say{What does ... do?}). Sometimes, however, the intended aspect was explicitly provided. We observed high-level, abstract aspects such as the purpose, rationale, or outcome of code, and low-level aspects such as the value of an expression, the effect of modifying a variable, or a trace of multiple lines of code. There was a slight difference in the proportion of interactions that contained an explicit intent between the two agents: 30\% for the control agent and 25\% for \approach{}. This difference was also observed between participants using and not using hypotheses: 29\% and 26\%, respectively.

\noindent\textbf{Conversation.} Some of the inputs exhibited additional conversational features, such as following up on a question posed to the agent's response or polite phrasing (e.g. \say{Can you explain ...?}). There was a noticeable difference in the prevalence of follow-up questions: 15\% of interactions for participants who did use hypotheses and 10\% for those who did not, and 15\% for the control approach compared to 11\% for \approach{}.

\noindent\textbf{Instruction.} In some instances, participants provided instructions on how the response should be formatted, such as asking for a detailed, short, step-by-step, or summarized response. This happened slightly more often for \approach{} than for the control approach: 5\% compared to 2\% of interactions, respectively. There was a negligible difference when looking at the participants grouped by their use (or not) of hypotheses.


\subsection{Code understanding (RQ2)}
\begin{table}[htbp]
    \scalefont{0.8}
    \sisetup{round-mode=places, round-precision=2}
    \centering
    \begingroup
        \setlength{\tabcolsep}{3pt}
        \begin{tabular}{l S[table-format=-1.2]S[table-format=1.2]S[table-format=1.2, table-space-text-post=\textsuperscript{***}] S[table-format=-3.2]S[table-format=3.2]S[table-format=1.2, table-space-text-post=\textsuperscript{***}]}
            \toprule
            & \multicolumn{3}{c}{Quiz score} 
            & \multicolumn{3}{c}{Quiz time (s)} \\
            & \multicolumn{3}{c}{$N{=}28$, $R^2_{\textrm{CS}}{=}0.15$} 
            & \multicolumn{3}{c}{$N{=}28$, $R^2_{\textrm{CS}}{=}0.44$} \\
    
            \cmidrule(lr){2-4}\cmidrule(lr){5-7}
    
            & \multicolumn{1}{c}{coeff.} 
            & \multicolumn{1}{c}{std}
            & \multicolumn{1}{c}{$p$-value}
            & \multicolumn{1}{c}{coeff.} 
            & \multicolumn{1}{c}{std}
            & \multicolumn{1}{c}{$p$-value} \\
            \midrule
    
            Intercept
            & 1.1857 & 	0.273 & 0.000\textsuperscript{***}
            & 593.1508 & 110.156 & 0.000\textsuperscript{***} \\
    
            Uses \approach{} 
            & 0.0526 & 0.102 & 0.606
            & -28.9396 & 43.241 & 0.503 \\
    
            Programm. exp. 
            & 0.0992 & 0.033 & 0.003\textsuperscript{**}
            & -39.3783 & 12.397 & 0.001\textsuperscript{**} \\
    
            Domain exp. 
            & -0.0108 & 0.029 & 0.711
            & -9.4231 & 8.961 & 0.293 \\
    
            LLM familiarity 
            & -0.0948 & 0.039 & 0.015\textsuperscript{*}
            & 36.0551 & 25.905 & 0.164 \\
    
            \bottomrule
            
            \multicolumn{7}{r}{\textsuperscript{*}$p < 0.05$, \textsuperscript{**}$p < 0.01$, \textsuperscript{***}$p < 0.001$}
        \end{tabular}
    \endgroup

    \captiondesc{Regression models showing the effect of several variables on quiz scores and completion times.}{The number of observations (2 per participant) is indicated with $N$.}
    \label{table:performance_regression}
\end{table}

\autoref{table:performance_regression} shows that the only factors found significantly to impact code understanding were programming experience and familiarity with LLMs. The positive coefficient for the scores model and negative coefficient for the time model shows that greater programming experience enables participants to score higher and finish the quizzes faster. Curiously, the reverse is true for familiarity with LLMs, although insignificantly so for quiz completion time.

\begin{table}[htbp]
    \scalefont{0.8}
    \sisetup{round-mode=places, round-precision=2}
    \centering
    \begingroup
        \setlength{\tabcolsep}{3pt}
        \begin{tabular}{l *{2}{S[table-format=-1.2]S[table-format=1.2]S[table-format=1.2, table-space-text-post=\textsuperscript{***}]}}
            \toprule
            & \multicolumn{3}{c}{No hypotheses} 
            & \multicolumn{3}{c}{Hypotheses} \\
            & \multicolumn{3}{c}{$N{=}16$, $R^2_{\textrm{CS}}{=}0.26$} 
            & \multicolumn{3}{c}{$N{=}12$, $R^2_{\textrm{CS}}{=}0.19$} \\
    
            \cmidrule(lr){2-4}\cmidrule(lr){5-7}
    
            & \multicolumn{1}{c}{coeff.} 
            & \multicolumn{1}{c}{std}
            & \multicolumn{1}{c}{$p$-value}
            & \multicolumn{1}{c}{coeff.} 
            & \multicolumn{1}{c}{std}
            & \multicolumn{1}{c}{$p$-value} \\
            \midrule
    
            Intercept
            & 0.9292 & 	0.273 & 0.001\textsuperscript{***}
            & 1.6550 & 0.364 & 0.000\textsuperscript{***} \\
    
            Uses \approach{} 
            & 0.3368 & 0.105 & 0.001\textsuperscript{**}
            & -0.2821 & 0.156 & 0.071 \\
    
            Programm. exp. 
            & 0.1235 & 0.038 & 0.001\textsuperscript{**}
            & 0.0624 & 0.032 & 0.052 \\
    
            Domain exp. 
            & -0.0473 & 0.027 & 0.082
            & 0.0074 & 0.030 & 0.802 \\
    
            LLM familiarity 
            & -0.0600 & 0.043 & 0.167
            & -0.1526 & 0.091 & 0.093 \\
    
            \bottomrule
            
            \multicolumn{7}{r}{\textsuperscript{*}$p < 0.05$, \textsuperscript{**}$p < 0.01$, \textsuperscript{***}$p < 0.001$}
        \end{tabular}
    \endgroup

    \captiondesc{Regression models showing the effect of several variables on quiz scores, wherein participants are grouped by their interaction style.}{The number of observations (2 per participant) is indicated with $N$.}
    \label{table:performance_regression_interaction}
\end{table}
No significant effects of using \approach{} are found when taking all participants into account. However, when fitting separate regression models to participants based on their interaction style, we do observe an effect (\autoref{table:performance_regression_interaction}). Using \approach{} contributes greatly to quiz scores for participants who did not phrase questions as hypotheses, but negatively impacts the scores of participants who did. The latter result is not significant, although it approaches the significance threshold of $p < 0.05$.


\subsection{Perceptions (RQ3)}

\noindent\textbf{Metrics.}
No significant differences in TAM and TLX scores were found between the two approaches (using two-sided Wilcoxon signed-rank tests, \autoref{table:perception_scores}). What stands out, however, is a lower mean but large standard deviation for the ``effort'' item of the TLX scores for \approach{} ($\text{M}{=}2.77$, $\text{SD}{=}4.46$) compared to the control agent ($\text{M}{=}4.00$, $\text{SD}{=}1.68$). This may suggest that \approach{} reduced perceived effort compared to the control approach, but this effect greatly varied per participant.

\begin{table}[htbp]
    \scalefont{0.8}
    \centering
    \begin{tabular}{l *{2}{S[table-format=-1.2, round-mode=places, round-precision=2]S[table-format=1.2, round-mode=places, round-precision=2]} S[table-format=1.2, round-mode=places, round-precision=2]}
        \toprule
        & \multicolumn{2}{c}{Control} 
        & \multicolumn{2}{c}{\approach{}} 
        &  \\
        \cmidrule(lr){2-3}\cmidrule(lr){4-5}

        & \multicolumn{1}{c}{mean}
        & \multicolumn{1}{c}{std}
        & \multicolumn{1}{c}{mean}
        & \multicolumn{1}{c}{std}
        & \multicolumn{1}{c}{$p$-value} \\
        \midrule

        Perceived usefulness 
        & 6.179487 & 0.759086        
        & 6.089744 & 0.759555
        & 0.6553647804541406 \\

        Perceived ease of use 
        & 6.269231 & 0.731369
        & 6.051282 & 0.967970
        & 0.5394981968180632 \\

        TLX score (avg.)
        & 0.448718 & 1.692008
        & 0.384615 & 2.582334 
        & 0.7353515625 \\

        \bottomrule
    \end{tabular}

    \caption{Comparison of TAM and TLX scores.}
    \label{table:perception_scores}
\end{table}


\noindent\textbf{Feedback.}
Both agents were seen as helpful and knowledgeable. Participants indicated that they liked the way their questions were answered and related back to the code. \approach{} was reported to understand the participants' questions well and produce clear answers ($n{=}3$ and $4$, respectively). This sentiment was also shared, although slightly less frequently, for the control approach ($n{=}2$ and $3$, respectively). Furthermore, P11 and P14 noted that they had to rephrase their question or provide more details to the control agent after it did not understand their question on the first try. P6 and P7 reported not always understanding the control agents' responses, causing P6 to ask for clarification several times.

\begin{figure}[htbp]
    \centering
    \setlength{\belowcaptionskip}{-10pt}
    \includegraphics[width=0.8\linewidth]{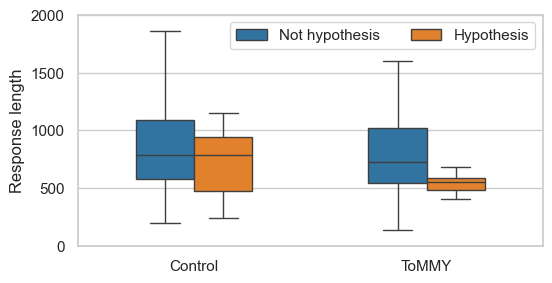}
    \caption{Distribution of response length, broken down by model and whether or not the question was phrased as a hypothesis.}
    \label{figure:response_length_model}
\end{figure}

Both the control approach and \approach{} were found to respond thoroughly, and in some cases with too much detail. Looking at \autoref{figure:response_length_model}, \approach{} seems to produce slightly shorter responses. However, when questions are phrased as hypotheses, \approach{} much more consistently replies with a shorter answer than the control approach. This suggests that \approach{} is better at recognizing when users just want to validate whether their interpretation of the code snippet is correct. Curiously, P1 found that the control approach appeared to \say{\textit{better fit the length of the answer to the question}}, which does not align with our interpretation of \autoref{figure:response_length_model}. Another counter-intuitive finding is P6 perceiving the control agent to produce shorter responses, while P4 perceived the opposite. Occasionally, the control agent did not answer as extensively as participants would have liked (P6, P10), a problem that was not encountered with \approach{}.

P4 noticed that \approach{} often seemed to respond with single paragraphs, whereas the control usually formatted its responses by emphasizing certain words. While the participant perceived \approach{}'s answers to be much shorter, they preferred the control agent's longer, formatted responses. This was echoed by P13, reporting that \approach{} alternated between formatting its responses as lists and as single paragraphs, which they perceived as inconsistent. Two more participants (P8, P11) commented positively on the format of the control approach's responses. Additionally, P3 and P4 noted
that \approach{} tended to end its answers with \say{\textit{if you have more questions, feel free to ask}}, which they experienced to be repetitive and useless.


\section{Discussion}
Several participants interacted with the agents by phrasing hypotheses, often aimed at fine-grained code features, and asking many follow-up questions. This aligns with Brook's top-down model of code comprehension~\cite{brooks1983}, which poses that programmers iteratively generate, test, and refine hypotheses by scanning the code. The other participants asked only open questions, sometimes phrased as instructions, and relatively frequently aimed these at the entire snippet. This corresponds to the bottom-up model of code comprehension~\cite{shneiderman1979}, stating that programmers recursively answer open ``what'' questions~\cite{letovsky1987} about chunks of code and combine the resulting knowledge until the entire program is understood. These interaction styles were not associated with differences in background experience.

Participants more often reported \approach{} to understand their question and produce clear answers than the control agent. Additionally, some participants commented negatively about the control agent on these aspects. Combined with participants less often explicitly stating their intent or asking follow-up questions to \approach, this suggests that \approach{} was able to better recognize participants' intent and adapt responses to their background experience. \approach{} also adjusted the length of its responses to the type of question more than the control agent did. However, this was not always perceived as such by participants. We suspect this is due to \approach{}'s responses often being formatted as a single paragraph, which might introduce usability issues seen in the literature~\cite{Liang2024}.

No differences in perceived usefulness, ease of use, and cognitive load were found between the two agents. Using \approach{} had a positive impact on quiz scores for participants with a bottom-up comprehension style, and a negative effect for participants exhibiting a top-down comprehension style. The latter group of participants more frequently asked targeted questions with explicit intent, which may have diminished the benefit of \approach{}'s mental state inference. For all of these metrics, a negative impact of \approach{}'s response format may have obscured other positive effects.

Our results imply LLMs can personalize responses independently on some aspects, but may need guidance on others. This guidance could entail explicitly prompting the LLM for certain elements of mental state ("What preferences does the user have regarding the response format?"), or even providing explicit instructions ("If the user asks a yes-no question, provide a minimal response.").

Future research include retaining and building on knowledge inferred in earlier interactions with \approach{}, to provide more robust and accurate mental inference. Retrieval-Augmented Generation~\cite{lewis2020} has shown promising results for storing user's mental data~\cite{leer2023}, and may be suitable for conversational programming assistants as well. Additionally, \approach{} being situated in a social, conversational environment allows for a more robust evaluation of new ToM prompting techniques than commonly-used basic tasks~\cite{ma2023}. Also, designing an interface for users to inspect and provide feedback on the inferred mental data would bring \approach{} closer to allowing a ``Mutual Theory of Mind''~\cite{wang2024}. The resulting increase in transparency and fairness towards the user would also better align the agent with existing human-AI interaction guidelines~\cite{amershi2019}.

\section{Limitations}
Prompt engineering and validation are notably difficult. We validated our prompt design choices through manual observations and a small-scale pilot study. However, small changes to \approach{}'s prompts may lead to unexpected side effects and ultimately different results.
Moreover, our sample was composed of students. 
Although they are novices \cite{steinmacher2016}, we acknowledge that additional research is necessary to consider the perspective of practitioners experienced with LLMs.
Also, within-subject studies are vulnerable to learning effects, fatigue, and other order effects. Varying the order in which participants interacted with the agents counteracted this.
While the code snippets were carefully selected and modified to have similar complexity, using other snippets may produce different outcomes.
Since we leverage qualitative research methods to categorize the open-ended questions asked in our surveys and the participant's interactions with the agents, we may have introduced categorization bias. To mitigate this bias, we conducted this process in pairs and carefully discussed categorization among the authors.

\section{Conclusion}
In this paper, we took the first steps towards personalizing developers' interactions with a code understanding assistant by proposing \approach{}, an assistant that can infer developers' mental states and adapt its responses. Although some novice programmers reported a good experience when using \approach{}, more research is still needed to understand how to structure its content. Interestingly, we also found that novices' interaction styles with the agent can significantly impact \approach{}'s effectiveness. Researchers and tool builders can leverage our approach to better adjust to users' needs, ensuring less cognitive load from users when interacting with the tool.

\textbf{Acknowledgments.} We thank the novice programmers who spent their time participating in our research.

\balance
\bibliographystyle{IEEEtran}
\bibliography{mairieli,jonan}

\end{document}